\documentclass[a4paper,12pt]{article}

\usepackage[utf8]{inputenc}

\usepackage{graphicx} 

\usepackage{subfig} 
\usepackage{amsthm,amsmath,amssymb}

\usepackage{hyperref}
\usepackage{authblk}

\usepackage{comment}

\usepackage[dvipsnames]{xcolor}

\title{Dyonic Einstein-Maxwell-scalar black holes:\\
the cold, the hot and the plunge}

\author[1]{Shun Chen
\thanks{\href{mailto:chuanyechen1016@gmail.com}
{chuanyechen1016@gmail.com}}}
\author{Xiao Yan Chew
\thanks{\href{mailto:xiao.yan.chew@just.edu.cn}
{xiao.yan.chew@just.edu.cn}}}
\author[2]{Jutta Kunz
\thanks{\href{mailto:jutta.kunz@uni-oldenburg.de}
{jutta.kunz@uni-oldenburg.de}}} 

\affil[1]{School of Science, Jiangsu University of Science and Technology, 212100, Zhenjiang, China}
\affil[2]{Institute of Physics, University of Oldenburg, Mailbox 2503,
D-26111 Oldenburg, Germany}

\date{\today}

\begin{document}

\maketitle
 
\begin{abstract}
We investigate dyonic nonlinearly scalarized black holes in Einstein-Maxwell-scalar theory.
The domain of existence of scalarized dyonic black holes consists of three branches.
The cold branch and the hot branch bifurcate at a minimal value of the charge, analogous to the purely electrically charged scalarized black holes. 
However, the presence of both charges allows for regular extremal black holes, leading to a third branch featuring a sudden plunge in Hawking temperature.
In fact, the presence of both electromagnetic charges introduces a factor $\Delta(\phi)$ in the source term of scalar field equations that vanishes when the coupling function $f(\phi)$ equals the ratio of the charges for some value of the scalar field $\phi_c$.
The scalar field of extremal black holes assumes precisely this value at the horizon, $\phi_H=\phi_c$. 
We demonstrate the plunge for the coupling function $f(\phi)=\exp(\alpha \phi^3)$.
\end{abstract}

\section{Introduction}

While discovered long ago in connection with neutron stars \cite{Damour:1993hw}, the phenomenon of spontaneous scalarization has received much attention during the last decade (see the review \cite{Doneva:2022ewd}), after the realization that black holes could be spontaneously scalarized as well \cite{Doneva:2017bvd,Silva:2017uqg,Antoniou:2017acq}.
Instead of being matter-induced as in the case of neutron stars, however, for black holes the scalarization is based on different mechanisms.
Coupling the scalar field with an appropriate coupling function to a curvature invariant like the Gauss-Bonnet term then produces curvature-induced spontaneous scalarization 
\cite{Doneva:2017bvd,Silva:2017uqg,Antoniou:2017acq,Antoniou:2017hxj,
Blazquez-Salcedo:2018jnn,Myung:2018iyq,Silva:2018qhn,Macedo:2019sem,Cunha:2019dwb,
Collodel:2019kkx,Macedo:2020tbm,Blazquez-Salcedo:2020rhf,Blazquez-Salcedo:2020caw,Dima:2020yac,Hod:2020jjy,Doneva:2020nbb,Herdeiro:2020wei,Berti:2020kgk}.
As noted subsequently, spontaneous scalarization of black holes can also be charged-induced when a scalar field is coupled with a suitable coupling function to the Maxwell invariant \cite{Herdeiro:2018wub,Myung:2018vug,Myung:2018jvi,Fernandes:2019rez,Brihaye:2019kvj,Myung:2019oua,Astefanesei:2019pfq,Zou:2019bpt,Zou:2020zxq,Lai:2022ppn,Guo:2023mda,Cheng:2025hdw}.

The coupling function $f(\phi)$ of the scalar field $\phi$ to the respective invariant determines the properties of the scalarized black holes.
In spontaneous scalarization, the black holes of General Relativity (GR) remain solutions of the field equations when the scalar field vanishes, $\phi=0$, and the first derivative of the coupling function with respect to the scalar field $\frac{df(\phi)}{d\phi}$ vanishes as well at $\phi=0$.
When the second derivative does not vanish, a tachyonic instability of the unscalarized GR solutions may arise and give rise to scalarized black holes.
In contrast, when the second derivative does vanish, the scalarized black holes are disconnected from the GR black holes.
Nonlinearly scalarized black holes have been studied both for coupling to the Gauss-Bonnet invariant \cite{Doneva:2021tvn,Blazquez-Salcedo:2022omw,Doneva:2022yqu,Lai:2023gwe,Zhang:2023jei,Pombo:2023lxg}
and for coupling to the Maxwell invariant \cite{Blazquez-Salcedo:2020nhs,LuisBlazquez-Salcedo:2020rqp,Blazquez-Salcedo:2020crd,Zhang:2021nnn,Chen:2023eru,Belkhadria:2023ooc,Xiong:2023bpl,Zhuang:2025eal}.

Previous studies of nonlinearly scalarized black holes in Einstein-Maxwell-scalar theory focused on black holes carrying electric charge only.
For a given coupling constant $\alpha$, then two branches of scalarized black holes emerge, the cold branch and the hot branch, while the bald Reissner-Nordst\"om black holes remain solutions.
The cold branch follows closely the Reissner-Nordst\"om solutions, until it bifurcates with the hot branch, while the hot branch ends in a singular solution with vanishing horizon area.
Here we study nonlinearly scalarized dyonic black holes in Einstein-Maxwell-scalar theory.
As shown by Astefanesei \textit{et al.}~\cite{Astefanesei:2019pfq} though some properties of scalarized dyonic black holes are valid more generally.
In particular, these dyonic black holes possess an extremal limit, where the ratio of the electric and magnetic charges is identical to the value of the coupling function at the horizon.

After specifying the model in section 2 and recalling the equations of motion for static spherically symmetric dyonic black holes, we present in section 3 exact solutions with a constant scalar field $\phi_c$.
We note, that such solutions arise, when the source term for the scalar field vanishes.
Since the source term contains besides the factor $\frac{d f(\phi)}{d\phi}$ also a factor denoted $\Delta(\phi)$, the vanishing of each factor at some associated $\phi_c$ leads to a distinct exact solution.
For $\Delta(\phi)$ to vanish, however, both electromagnetic charges are necessary.
In section 4 we first recall some general properties of dyonic black holes, including their extremal limit, and then consider a specific example for the coupling function, $f(\phi)=\exp{(\alpha \phi^3)}$.
For this coupling function we then solve the field equations numerically.
We show that the presence of both charges leads to a drastic plunge in temperature as the extremal black hole is approached.
We conclude in section 5.

\section{Model}

The EMS family of models is defined by the following action (we set $c = G = 4\pi\varepsilon_0 = 1$):
\begin{equation}
S = \frac{1}{16\pi} \int d^4x \sqrt{-g}
\left(
R - 2\,\partial_\mu \phi\, \partial^\mu \phi
- f(\phi)\, F_{\mu\nu} F^{\mu\nu}
\right).
\label{action}
\end{equation}
Here $R$ is the Ricci scalar,
$F_{\mu\nu} = \partial_\mu A_\nu - \partial_\nu A_\mu$ is the Maxwell field strength tensor,
and $\phi$ is the scalar field.
The coupling function $f(\phi)$ governs the non-minimal coupling of $\phi$ to the electromagnetic field. 
After variation, we get the set of field equations
\begin{equation}
R_{\mu\nu} - \frac{1}{2} R g_{\mu\nu}
=
2 \left[
\partial_\mu \phi \, \partial_\nu \phi
- \frac{1}{2} g_{\mu\nu} \, \partial_\rho \phi \, \partial^\rho \phi
+ f(\phi)
\left(
F_{\mu\rho} F_{\nu}{}^{\rho}
- \frac{1}{4} g_{\mu\nu} F_{\rho\sigma} F^{\rho\sigma}
\right)
\right] ,
\label{eqmetric}
\end{equation}
\begin{equation}
\frac{1}{\sqrt{-g}} \, \partial_\mu
\left(
\sqrt{-g} \, \partial^\mu \phi
\right)
=
\frac{1}{4} \frac{d f(\phi)}{d \phi}
F_{\rho\sigma} F^{\rho\sigma} ,
\label{eqphi}
\end{equation}
\begin{equation}
\partial_\mu
\left(
\sqrt{-g} \, f(\phi) \, F^{\mu\nu}
\right)
= 0 .
\label{eqem}
\end{equation}

To obtain dyonic static spherically solutions we employ the following Ansatz for the metric
\begin{equation}
ds^2
=
- N(r)\, e^{-2\sigma(r)} dt^2
+ \frac{dr^2}{N(r)}
+ r^2 \left( d\theta^2 + \sin^2\theta \, d\varphi^2 \right) ,
\label{metric}
\end{equation}
with
\begin{equation}
N(r)=1 -\frac{m(r)}{r} .
\label{massfun}
\end{equation}
The corresponding Ansatz for the Maxwell field of a dyonic black hole reads
\begin{equation}
 A= V(r) dt+P \cos \theta d\varphi 
 \label{em}
\end{equation}
with the electric potential $V(r)$ and the magnetic charge $P$,
and the scalar field is given by $\phi(r)$.
The Maxwell equation (\ref{eqem}) then yields 
\cite{Astefanesei:2019pfq}
\begin{equation}
    V'(r) = \frac{e^{-\sigma}}{r^2} \frac{Q}{f(\phi)}
    \label{first} ,
\end{equation}
where $Q$ is the electric charge.
With this Ansatz the resulting set of equations becomes
\begin{equation}
m'(r)
=
\frac{1}{2} r^2 N(r)\, \phi'(r)^2
+ \frac{1}{2 r^2}
\left(
\frac{Q^2}{f(\phi)} + f(\phi)\, P^2
\right) ,
\label{eqmass}
\end{equation}
\begin{equation}
\sigma'(r) + r\, \phi'(r)^2 = 0 ,
\label{eqsigma}
\end{equation}
\begin{equation}
\left(
e^{-\sigma(r)} r^2 N(r)\, \phi'(r)
\right)'
+
\frac{e^{-\sigma(r)}}{2 r^2 f(\phi)}
\frac{d f(\phi)}{d \phi}
\left(
\frac{Q^2}{f(\phi)} - f(\phi)\, P^2
\right)
= 0 .
\label{eqphi}
\end{equation}
The equations possess electro-magnetic duality \cite{Astefanesei:2019pfq}
\begin{equation}
\label{dual}
\{P\to Q, \ \ Q\to P \}\ \ {\rm and} \ \ f(\phi)\to 1/f(\phi) \ .
\end{equation}
Taking both $Q$ and $P$ positive and $ Q\geq P$ limits their ratio to $P/Q \le 1$.

\section{Exact solutions}

As noted before \cite{Astefanesei:2019pfq}, the Reissner-Nordstr\"om solution is a  solution of the field equations for $f(\phi)=1$, and a constant scalar field $\phi=\phi_c$
\begin{eqnarray}
V(r) = - \frac{Q}{r}\ , 
\qquad N(r) = 1-\frac{2M}{r}+\frac{Q^2+P^2}{r^2}\ , 
\qquad \sigma(r) = 0 \ , 
\label{RN1}
\end{eqnarray}
where $M$ is the ADM mass of the dyonic black hole.
It remains a solution of the field equations if 
\begin{equation}
\left .\frac{d f(\phi)}{d\phi} \right|_{\phi_c} =0 ,
\label{eq_phic2}
\end{equation}
for $\phi(r)=\phi_c=0$, which presents the basic requisite of the phenomenon of scalarization of Reissner-Nordstr\"om black holes.

Another exact solution of the field equations is obtained, if eq.~(\ref{eq_phic2}) holds in addition for a constant scalar field $\phi(r)=\phi_c \ne 0$. 
The Einstein equations then reduce to
\begin{eqnarray}
m'(r)
& = &
\frac{1}{2r^2} \left (\frac{Q^2}{f(\phi_c)} + f(\phi_c)\, P^2
\right) ,
\label{eq_mc2}
\\
\sigma'(r) & = & 0 \ ,
\label{eq_sigc2}
\end{eqnarray}
with solution
\begin{equation}
m(r) = M - \frac{1}{2r} \left (\frac{Q^2}{f(\phi_c)} + f(\phi_c)\, P^2
\right)
\ , \ \ \ 
\sigma(r) = 0 \ .
\label{RN_c2}
\end{equation}

The source term of the scalar field vanishes, however, as well, when the last factor vanishes,
\begin{equation}
  \Delta(\phi) =  \left(
\frac{Q^2}{f(\phi)} - f(\phi)\, P^2
\right)
= 0 ,
\end{equation}
abbreviated by $\Delta(\phi)$.
Thus a further exact solution with a constant scalar field $\phi=\phi_c$ is found when $\phi_c$ is a solution of 
\begin{equation}
\Delta (\phi_c) = 0
\ \ \ \Longleftrightarrow \ \ \ 
f(\phi_c) = \frac{Q}{P} .
\label{eq_phic3}
\end{equation}
The Einstein equations then reduce to
\begin{eqnarray}
m'(r)
& = &
 \frac{P\, Q }{r^2} \ , 
\label{eq_mc}
\\
\sigma'(r) & = & 0 \ ,
\label{eq_sigc3}
\end{eqnarray}
with solution
\begin{equation}
m(r) = M - \frac{P\, Q }{r} \ , \ \ \ 
\sigma(r) = 0 \ .
\label{RN_c3}
\end{equation}
Thus, the extremal black hole satisfies
\begin{eqnarray}
\label{ex0}
f(\phi_H)=\frac{Q}{P}\ , \ \ \  r_H=\sqrt{2 PQ} \ .
\end{eqnarray}

In all these cases, the metric is asymptotically flat, but the 
scalar field 
is non-vanishing in the asymptotic region. 

\section{Scalarized black holes}

\subsection{General considerations}

The expansions at the horizon and at radial infinity for scalarized black holes have been presented in \cite{Astefanesei:2019pfq}.
Accordingly, a set of boundary conditions yielding non-extremal black holes that are regular on and outside their horizons, that are asymptotically flat, and possess a scalar field that is vanishing at infinity is given by
\begin{eqnarray}
    && \left. m(r)\right|_{r_H} = \frac{r_H}{2} \ , \ \ \ 
    \left. \phi(r)\right|_{r_H} = \phi_H \ , \ \ \ 
    \label{bc1} \\
     && \left. \phi'(r) \right|_{r_H} 
    =  \frac{1}{2r_H}  \frac{df(\phi)}{d\phi}\bigg |_{\phi_H} 
\frac{\left(\frac{Q^2}{f(\phi_H)} - f(\phi_H)P^2 \right)}
 {\left({Q^2}  -r_H^2 f(\phi_H) + f(\phi_H)^2P^2  \right)} \ , 
 \label{bc2} \\[2ex]
 &&\left. \sigma(r)\right|_\infty =0  \ , \ \ \
 \left. V(r)\right|_\infty =0  \ , \ \ \
 \left. \phi(r)\right|_\infty =0  \ .
 \label{bc3}
\end{eqnarray}
To be able to impose all boundary conditions, we introduce an auxiliary equation for the charge $Q$, $Q'(r)=0$.

As shown in \cite{Astefanesei:2019pfq}, for extremal scalarized black holes the field equations imply
\begin{eqnarray}
\label{ex1}
 Q =\frac{r_H\sqrt{f(\phi_H)}}{\sqrt{2}}\ , \ \ \ 
 P=\frac{r_H}{\sqrt{2f(\phi_H)}} \ , 
\end{eqnarray}
and therefore
\begin{eqnarray}
\label{ex2}
f(\phi_H)=\frac{Q}{P}\ , \ \ \  r_H=\sqrt{2 PQ} \ ,
\end{eqnarray}
showing immediately the necessity of a finite magnetic charge $P$ for a finite horizon area.
Moreover, we note that relations (\ref{ex2}) 
correspond to relations (\ref{ex0}) 
of that constant $\phi$ solution.

The asymptotics of the solutions determine their global charges.
The thermodynamics of the solutions is determined by their horizon properties.
The horizon area $A_H$ and the Hawking temperature $T_H$ are given by
\begin{equation}
  A_H=4\pi r_H^2 \ , \ \ \  T_H=\frac{1}{4\pi}N'(r_H)e^{-\delta_H} .
\end{equation}
Along with \cite{Astefanesei:2019pfq} we consider the dimensionless quantities
\begin{equation}
q = \frac{\sqrt{Q^2+P^2}}{M} \ , \ \ \
\beta = \frac{P}{Q} \ , \ \ \
a_H = \frac{A_H}{16\pi M^2} \ , \ \ \ 
t_H = 8\pi T_H M .
\end{equation}

\subsection{Specific example}

We now present a set of dyonic scalarized black holes choosing an exponential coupling function
\begin{equation}
    f(\phi) = e^{\alpha \phi^3} .
    \label{coupling}
\end{equation}
This coupling function satisfies
\begin{equation}
   \left. \frac{d f(\phi)}{d\phi} \right|_{\phi=0}=0 \ , \ \ \
   \left. \frac{d^2 f(\phi)}{d\phi^2} \right|_{\phi=0}=0 \ ,
\end{equation}
and therefore, leads to so-called nonlinearly scalarized black holes.

We solve the set of equations numerically, subject to the boundary conditions (\ref{bc1})-(\ref{bc3}).
In the calculations, we fix a value of the coupling constant $\alpha$ and impose a constant ratio $\beta=P/Q$.
To obtain the associated families of solutions, we retain a fixed horizon radius $r_H$ and slowly increase the horizon value of the scalar field $\phi_H$.
We then extract the physical quantities from the obtained sets of solutions.

 \begin{figure}[ht!]
 	\begin{center}
    \includegraphics[height=.34\textwidth, angle =0 ]{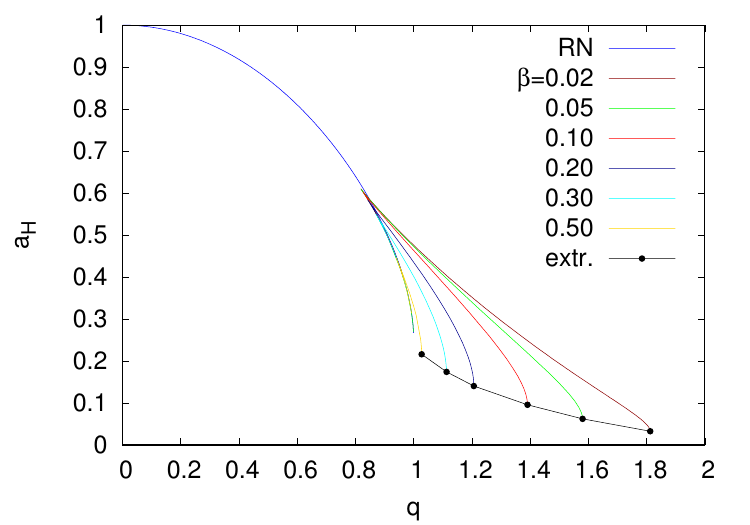}
    \includegraphics[height=.34\textwidth, angle =0 ]{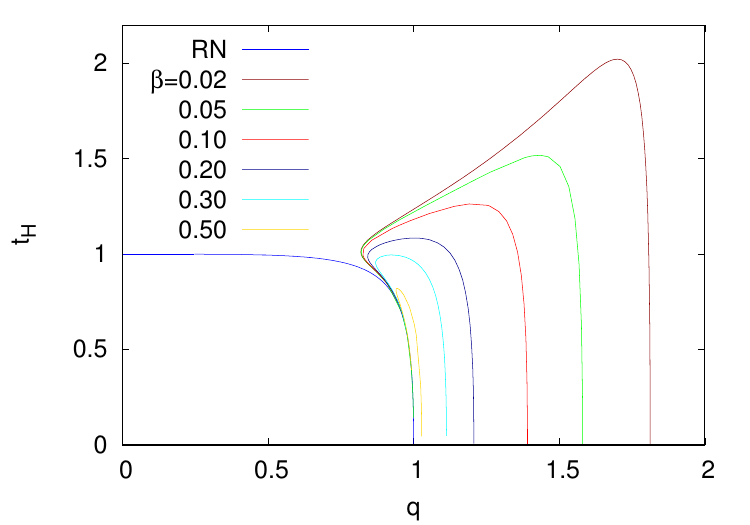} 
 	\end{center}
 	\caption{  
 		(a) The dimensionless horizon area $a_H$ is shown versus the charge parameter $q$. 
        (b) The dimensionless temperature $t_H$ is shown vs the charge parameter $q$.
 	}
 	\label{fig1}
 \end{figure}

Figure \ref{fig1}(a) shows the dimensionless horizon area $a_H$ versus the charge parameter $q$ for several sets of solutions for increasing charge ratio $\beta$.
Also shown is the set of Reissner-Nordstr\"om black holes.
The black dots mark the extremal endpoints of the scalarized solution sets.
Clearly, the extremal black holes possess a finite area for $\beta \ne 0$.

For $\beta=0$, the scalarized black holes possess a cold branch and a hot branch \cite{Blazquez-Salcedo:2020nhs}.
The cold branch closely follows the bald branch of Reissner-Nordstr\"om black holes, starting in the vicinity of the extremal Reissner-Nordstr\"om black hole \cite{Blazquez-Salcedo:2020nhs,LuisBlazquez-Salcedo:2020rqp,Blazquez-Salcedo:2020crd}.
At a minimal value of the charge parameter $q$, the cold branch bifurcates with the hot branch.
The hot branch then approaches a singular solution at a maximal value of $q$, that depends on the coupling constant $\alpha$.

For finite $\beta$, the hot branch experiences, however, a sudden dramatic plunge in temperature, as the scalarized solutions tend toward their regular extremal limit.
This is illustrated in Fig.~\ref{fig1}(b), where the dimensionless temperature $t_H$ is shown versus the charge parameter $q$.
This plunge gets the more dramatic the smaller the value of $\beta$.
The figure further shows that the domain of existence of scalarized solutions decreases with increasing values of $\beta$.

 \begin{figure}[ht!]
 	\begin{center}
    \includegraphics[height=.34\textwidth, angle =0 ]{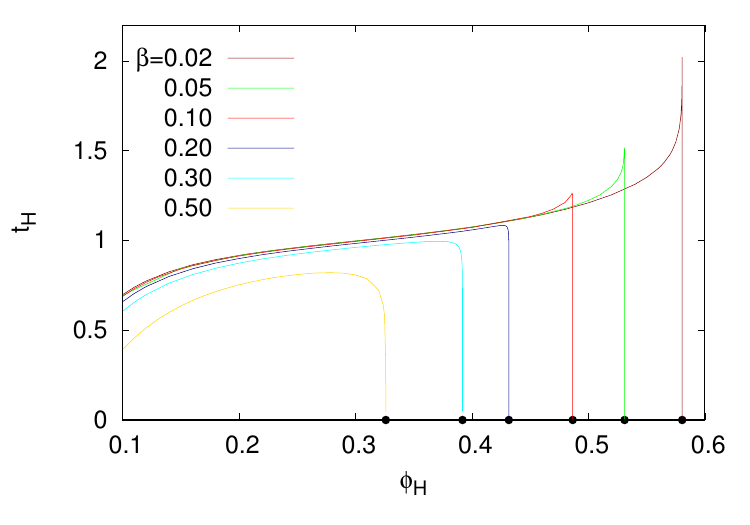}
    \includegraphics[height=.34\textwidth, angle =0 ]{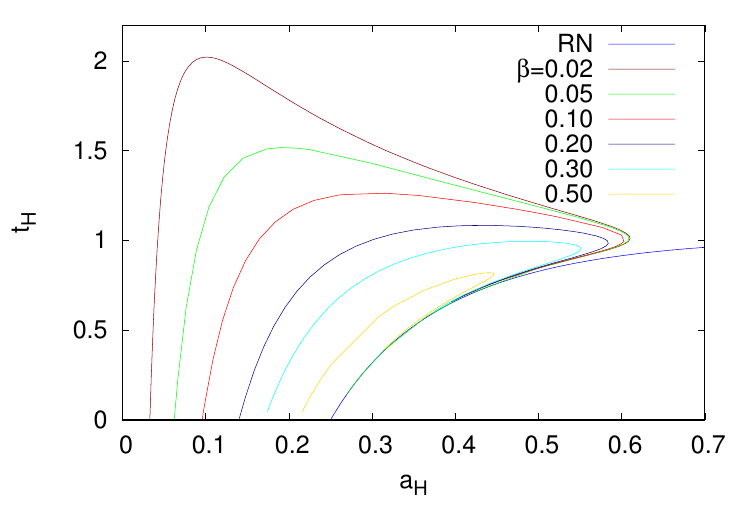} 
 	\end{center}
 	\caption{  
 		(a) The dimensionless temperature $t_H$ is shown vs the scalar field at the horizon $\phi_H$. 
        (b) The dimensionless temperature $t_H$ is shown vs the horizon area $a_H$.
 	}
 	\label{fig2}
 \end{figure}

 Figure \ref{fig2}(a) illustrates the plunge even more dramatically.
 Here we exhibit the dimensionless temperature $t_H$ versus the value of the scalar field at the horizon $\phi_H$ for the same sets of solutions.
 We note that the smooth dependence of $t_H$ on $\phi_H$ reaches a sudden end, as the horizon value of the scalar field approaches $\phi_H=\phi_c$, i.e., its value of the extremal black hole.
 Here a spike develops that gets the sharper the smaller $\beta$.
 Beyond the spike, the temperature plunges basically vertically to zero.\footnote{This plunge makes the scalar field horizon value $\phi_H$ inadequate for the numerical calculations, requiring another parameter to follow the plunge, e.g., $V_H$.}
 
\begin{figure}[ht!]
 	\begin{center}
    \includegraphics[height=.34\textwidth, angle =0 ]{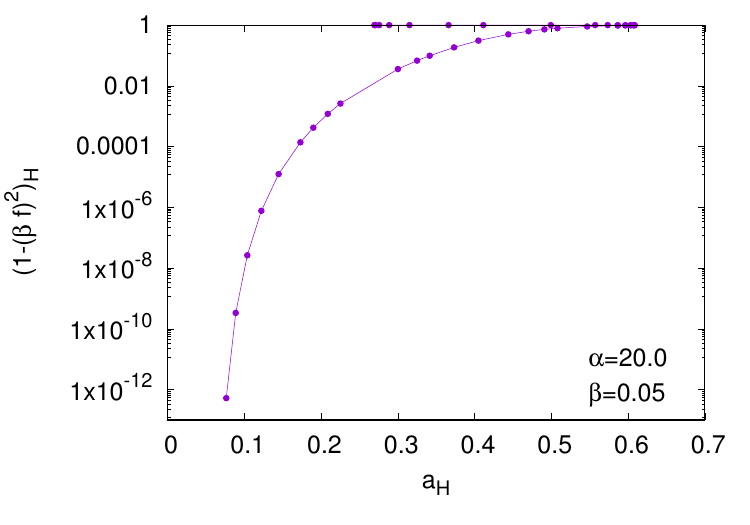}
    \includegraphics[height=.34\textwidth, angle =0 ]{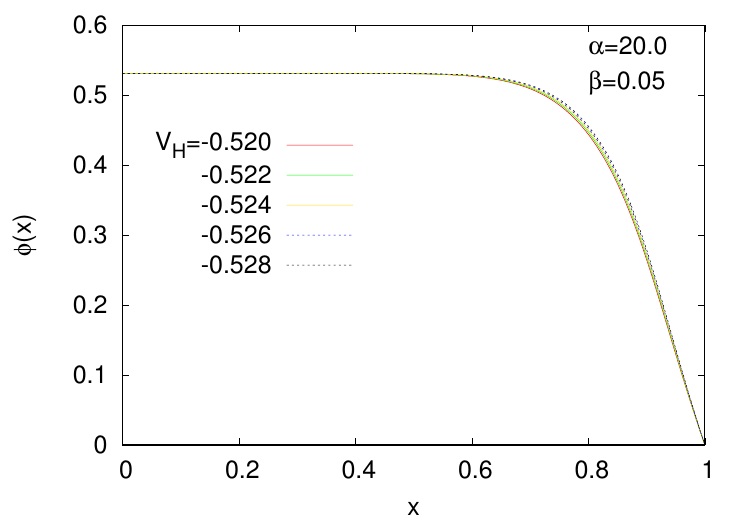} 
 	\end{center}
 	\caption{  
 		(a) The factor $(1- (\beta f(\phi_H))^2)$ of $\Delta(\phi_H)$ is shown on a logarithmic scale vs the dimensionless horizon area $a_H$ for $\alpha=20$, $\beta=0.05$. 
        (b) The scalar field $\phi(x)$ is shown vs the compactified dimensionless radial coordinate $x=1-\frac{r_H}{r}$ for the last few values of the electric potential at the horizon $V_H$ close to the extremal black hole solution for $\alpha=20$, $\beta=0.05$.
 	}
 	\label{fig3}
 \end{figure}

As the scalarized black hole solutions approach the extremal solution, the quantity $\Delta(\phi_H)$ rapidly approaches zero.
In the limit, the equations (\ref{ex1})-(\ref{ex2}) are therefore attained.
The telling factor in $\Delta(\phi_H)$ is $(1- (\beta f(\phi_H))^2)$, which vanishes for the extremal solution, see eq.~(\ref{ex2}). 
This factor is exhibited in Fig.~\ref{fig3}(a) for the family of solutions for $\alpha=20$, $\beta=0.05$.
The dots mark the respective values of the numerical calculations.

At the same time we note that the derivative of the scalar field at the horizon eq.~(\ref{bc2}) is proportional to $\Delta(\phi_H)$.
As the extremal black hole is approached, the scalar field changes very little close to the horizon.
This is to be expected since the next to leading order term in the expansion of the scalar field at the horizon of the extremal black hole is only of power $(r-r_H)^k$, where \cite{Astefanesei:2019pfq}
\begin{equation}
   k = \frac{1}{2}  \left( -1 + \sqrt{ 1 + 2 
   \left( \frac{f'(\phi_H)}{f(\phi_H)} \right)^2} \right) > 0 \ .
\label{exk}
\end{equation}
We exhibit the scalar field function $\phi(x)$ versus the compactified dimensionless coordinate $x=1-\frac{r_H}{r}$ (with $r_H=1$) employed in the numerical calculations in Fig.~\ref{fig3}(b) for the last few points of Fig.~\ref{fig3}(a) for $\alpha=20$, $\beta=0.05$.
With 
$\phi_H=0.53108$, in this example 
$k=11.47651$.

\section{Conclusions}

We have studied nonlinearly scalarized black holes in Einstein-Maxwell-scalar theory carrying both electric and magnetic charge.
In contrast to the solely electrically charged scalarized black holes, which possess a cold branch and a hot branch without a regular extremal limit, the dyonic scalarized black holes feature a regular extremal limit.
The extremal black holes satisfy the relation $f(\phi_H)=Q/P$ at their horizon.
This relation corresponds precisely to the relation necessary for the source term of the scalar field to vanish, $\Delta(\phi_c)=0$ for $\phi_c=\phi_H$.

To demonstrate the effect of the presence of both charges, we have chosen the coupling function $f(\phi)=\exp{(\alpha \phi^3)}$.
For this coupling function we have solved the field equations for fixed values of the coupling constant $\alpha$ and the charge ratio $\beta=P/Q$.
The resulting sets of solutions show a similar dependence of the horizon area on the charge as the singly charged solutions, except for their finite area endpoint formed by the extremal black hole.
The cold branch bifurcates with the hot branch at a minimal value of the charge.
However, when considering the temperature dependence, the hot branch ends suddenly, giving rise to a dramatic plunge to the extremal black hole.
In fact, the plunge is most dramatic when the scalar field at the horizon $\phi_H$ is considered for small values of the charge ratio $\beta$.

The next steps will be to address the stability and quasinormal modes of these nonlinearly scalarized dyonic black holes, and to construct their rotating generalizations.
Since the electromagnetic charge of astrophysical black holes is thought to be very small (see e.g., \cite{Zajacek:2018vsj,Zajacek:2018vsj,Zajacek:2018ycb,Allahyari:2019jqz,Pulice:2023dqw,Gu:2023eaa}),
the physical relevance of these solutions might reside in the dark sector.

\section*{Acknowledgement}
We gratefully acknowledge support by Burkhard Kleihaus.
SC is supported by the Research and Practice Innovation Plan for Graduate Students in Jiangsu Province (no: KYCX25-4317). XYC is supported by the starting grant of Jiangsu University of Science and Technology (JUST) and National Science Foundation of China (no: W2533026).

\begin{thebibliography}{99}

\bibitem{Damour:1993hw} 
T.~Damour and G.~Esposito-Farese,
Phys.\ Rev.\ Lett.\  {\bf 70}, 2220 (1993)

\bibitem{Doneva:2022ewd}
D.~D.~Doneva, F.~M.~Ramazano\u{g}lu, H.~O.~Silva, T.~P.~Sotiriou and S.~S.~Yazadjiev,
Rev. Mod. Phys. \textbf{96}, 015004 (2024)


\bibitem{Doneva:2017bvd} 
D.~D.~Doneva and S.~S.~Yazadjiev,
Phys.\ Rev.\ Lett.\  {\bf 120}, 131103 (2018)

\bibitem{Silva:2017uqg} 
H.~O.~Silva, J.~Sakstein, L.~Gualtieri, T.~P.~Sotiriou and E.~Berti,
Phys.\ Rev.\ Lett.\  {\bf 120}, 131104 (2018)

\bibitem{Antoniou:2017acq} 
G.~Antoniou, A.~Bakopoulos and P.~Kanti,
Phys.\ Rev.\ Lett.\  {\bf 120}, 131102 (2018)


\bibitem{Antoniou:2017hxj} 
  G.~Antoniou, A.~Bakopoulos and P.~Kanti,
  Phys.\ Rev.\ D {\bf 97}, 084037 (2018)

\bibitem{Blazquez-Salcedo:2018jnn} 
  J.~L.~Blázquez-Salcedo, D.~D.~Doneva, J.~Kunz and S.~S.~Yazadjiev,
  Phys.\ Rev.\ D {\bf 98}, 084011 (2018)

\bibitem{Myung:2018iyq}
Y.~S.~Myung and D.~C.~Zou, 
Phys.\ Rev.\ D {\bf 98}, 024030 (2018) 

\bibitem{Silva:2018qhn} 
  H.~O.~Silva, C.~F.~B.~Macedo, T.~P.~Sotiriou, L.~Gualtieri, J.~Sakstein and E.~Berti,
  Phys.\ Rev.\ D {\bf 99}, 064011 (2019)

\bibitem{Macedo:2019sem} 
  C.~F.~B.~Macedo, J.~Sakstein, E.~Berti, L.~Gualtieri, H.~O.~Silva and T.~P.~Sotiriou,
  Phys.\ Rev.\ D {\bf 99}, 104041 (2019)

\bibitem{Cunha:2019dwb} 
  P.~V.~P.~Cunha, C.~A.~R.~Herdeiro and E.~Radu,
  Phys.\ Rev.\ Lett.\  {\bf 123}, 011101 (2019)

\bibitem{Collodel:2019kkx}
L.~G.~Collodel, B.~Kleihaus, J.~Kunz and E.~Berti,
Class. Quant. Grav. \textbf{37}, 075018 (2020)

\bibitem{Macedo:2020tbm}
C.~F.~B.~Macedo,
Int. J. Mod. Phys. D \textbf{29}, 2041006 (2020)

\bibitem{Blazquez-Salcedo:2020rhf}
J.~L.~Bl\'azquez-Salcedo, D.~D.~Doneva, S.~Kahlen, J.~Kunz, P.~Nedkova and S.~S.~Yazadjiev,
Phys. Rev. D \textbf{101}, 104006 (2020)

\bibitem{Blazquez-Salcedo:2020caw}
J.~L.~Bl\'azquez-Salcedo, D.~D.~Doneva, S.~Kahlen, J.~Kunz, P.~Nedkova and S.~S.~Yazadjiev,
Phys. Rev. D \textbf{102}, 024086 (2020)

\bibitem{Dima:2020yac}
A.~Dima, E.~Barausse, N.~Franchini and T.~P.~Sotiriou,
Phys. Rev. Lett. \textbf{125}, 231101 (2020)

\bibitem{Hod:2020jjy}
S.~Hod,
Phys. Rev. D \textbf{102}, 084060 (2020)

\bibitem{Doneva:2020nbb}
D.~D.~Doneva, L.~G.~Collodel, C.~J.~Kr\"uger and S.~S.~Yazadjiev,
Phys. Rev. D \textbf{102}, 104027 (2020)

\bibitem{Herdeiro:2020wei}
C.~A.~R.~Herdeiro, E.~Radu, H.~O.~Silva, T.~P.~Sotiriou and N.~Yunes,
Phys. Rev. Lett. \textbf{126}, 011103 (2021)

\bibitem{Berti:2020kgk}
E.~Berti, L.~G.~Collodel, B.~Kleihaus and J.~Kunz,
Phys. Rev. Lett. \textbf{126}, 011104 (2021)



\bibitem{Herdeiro:2018wub}
C.~A.~R.~Herdeiro, E.~Radu, N.~Sanchis-Gual and J.~A.~Font,
Phys. Rev. Lett. \textbf{121}, 101102 (2018)

\bibitem{Myung:2018vug}
  Y.~S.~Myung and D.~C.~Zou,
  Eur.\ Phys.\ J.\ C {\bf 79}, 273 (2019)

\bibitem{Myung:2018jvi}
Y.~S.~Myung and D.~C.~Zou,
Phys. Lett. B \textbf{790}, 400 (2019)

\bibitem{Fernandes:2019rez}
P.~G.~S.~Fernandes, C.~A.~R.~Herdeiro, A.~M.~Pombo, E.~Radu and N.~Sanchis-Gual,
Class. Quant. Grav. \textbf{36}, 134002 (2019) 
[erratum: Class. Quant. Grav. \textbf{37}, 049501 (2020)]

\bibitem{Brihaye:2019kvj}
Y.~Brihaye and B.~Hartmann,
Phys. Lett. B \textbf{792}, 244 (2019)

\bibitem{Myung:2019oua}
Y.~S.~Myung and D.~C.~Zou,
Eur. Phys. J. C \textbf{79}, 641 (2019)

\bibitem{Astefanesei:2019pfq}
D.~Astefanesei, C.~Herdeiro, A.~Pombo and E.~Radu,
JHEP \textbf{10}, 078 (2019)

\bibitem{Zou:2019bpt}
D.~C.~Zou and Y.~S.~Myung,
Phys. Rev. D \textbf{100}, 124055 (2019)
  
\bibitem{Zou:2020zxq}
D.~C.~Zou and Y.~S.~Myung,
  Phys.\ Rev.\ D {\bf 102}, 064011 (2020)

\bibitem{Lai:2022ppn}
M.~Y.~Lai, Y.~S.~Myung, R.~H.~Yue and D.~C.~Zou,
Phys. Rev. D \textbf{106}, 084043 (2022)

\bibitem{Guo:2023mda}
G.~Guo, P.~Wang, H.~Wu and H.~Yang,
JHEP \textbf{10}, 076 (2023)

\bibitem{Cheng:2025hdw}
L.~Cheng, G.~Guo, P.~Wang and H.~Yang,
[arXiv:2506.01773 [gr-qc]]






\bibitem{Doneva:2021tvn}
D.~D.~Doneva and S.~S.~Yazadjiev,
\textit{Phys. Rev. D} \textbf{105}, L041502 (2022)

\bibitem{Blazquez-Salcedo:2022omw}
J.~L.~Bl\'azquez-Salcedo, D.~D.~Doneva, J.~Kunz and S.~S.~Yazadjiev,
\textit{Phys. Rev. D} \textbf{105}, 124005 (2022)

\bibitem{Doneva:2022yqu}
D.~D.~Doneva, L.~G.~Collodel and S.~S.~Yazadjiev,
\textit{Phys. Rev. D} \textbf{106}, 104027 (2022) 

\bibitem{Lai:2023gwe}
M.~Y.~Lai, D.~C.~Zou, R.~H.~Yue and Y.~S.~Myung,
Phys. Rev. D \textbf{108}, 084007 (2023)

\bibitem{Zhang:2023jei}
S.~J.~Zhang,
Eur. Phys. J. C \textbf{83}, 950 (2023)

\bibitem{Pombo:2023lxg}
A.~M.~Pombo and D.~D.~Doneva,
Phys. Rev. D \textbf{108}, 124068 (2023)




\bibitem{Blazquez-Salcedo:2020nhs}
J.~L.~Bl{\'a}zquez-Salcedo, C.~A.~R.~Herdeiro, J.~Kunz, A.~M.~Pombo and E.~Radu,
Phys. Lett. B \textbf{806}, 135493 (2020)

\bibitem{LuisBlazquez-Salcedo:2020rqp}
J.~Luis Bl{\'a}zquez-Salcedo, C.~A.~R.~Herdeiro, S.~Kahlen, J.~Kunz, A.~M.~Pombo and E.~Radu,
Eur. Phys. J. C \textbf{81}, 155 (2021)

\bibitem{Blazquez-Salcedo:2020crd}
J.~L.~Bl{\'a}zquez-Salcedo, S.~Kahlen and J.~Kunz,
Symmetry \textbf{12}, 2057 (2020)

\bibitem{Zhang:2021nnn}
C.~Y.~Zhang, Q.~Chen, Y.~Liu, W.~K.~Luo, Y.~Tian and B.~Wang,
Phys. Rev. Lett. \textbf{128},  161105,(2022)

\bibitem{Chen:2023eru}
Q.~Chen, Z.~Ning, Y.~Tian, B.~Wang and C.~Y.~Zhang,
Phys. Rev. D \textbf{108}, 084016 (2023)

\bibitem{Belkhadria:2023ooc}
Z.~Belkhadria and A.~M.~Pombo,
Phys. Rev. D \textbf{110}, 044014 (2024)

\bibitem{Xiong:2023bpl}
W.~Xiong, C.~Y.~Zhang and P.~C.~Li,
JCAP \textbf{09}, 031 (2024)






\bibitem{Zhuang:2025eal}
Z.~Zhuang, K.~Meng and H.~Zhang,
[arXiv:2505.22033 [gr-qc]]


\bibitem{Zajacek:2018vsj}
M.~Zaja{\v{c}}ek, A.~Tursunov, A.~Eckart, S.~Britzen, E.~Hackmann, V.~Karas, Z.~Stuchl{\'\i}k, B.~Czerny and J.~A.~Zensus,
J. Phys. Conf. Ser. \textbf{1258}, 012031 (2019)

\bibitem{Zajacek:2018ycb}
M.~Zaja{\v{c}}ek, A.~Tursunov, A.~Eckart and S.~Britzen,
Mon. Not. Roy. Astron. Soc. \textbf{480}, 4408 (2018)

\bibitem{Allahyari:2019jqz}
A.~Allahyari, M.~Khodadi, S.~Vagnozzi and D.~F.~Mota,
JCAP \textbf{02}, 003 (2020)

\bibitem{Pulice:2023dqw}
B.~Puli{\c{c}}e, R.~C.~Pantig, A.~{\"O}vg{\"u}n and D.~Demir,
Class. Quant. Grav. \textbf{40}, 195003 (2023)

\bibitem{Gu:2023eaa}
H.~P.~Gu, H.~T.~Wang and L.~Shao,
Phys. Rev. D \textbf{109}, 024058 (2024)




\end{thebibliography}
\end{document}